\documentclass[12pt,a4paper]{article}
\usepackage{color}
\usepackage{graphicx}
\usepackage{float}
\usepackage[margin=2cm]{geometry}
\usepackage{amsmath}
\usepackage{tikz}
\usepackage{booktabs}
\usepackage{amssymb}
\usepackage[font={small,it}]{caption}
\usepackage{tikz}
\usepackage{titletoc}
\usepackage{mathtools}
\usepackage{braket}
\usepackage{subcaption}
\usepackage{subfloat}
\usepackage{cleveref}
\usepackage[utf8]{inputenc}
\usepackage[T1]{fontenc}
\usepackage[english]{babel}
\usepackage{verbatim}
\usepackage{tabu}
\usepackage{cite}
\usetikzlibrary{decorations.pathmorphing}
\usetikzlibrary{arrows,decorations.markings}
\usetikzlibrary{shapes}
\usepackage{wrapfig}
\usepackage{animate}
\usepackage[space]{grffile}
\usepackage[labelfont=bf]{caption}
\usepackage{upgreek}
\usepackage{authblk}
\usepackage[separate-uncertainty = true,multi-part-units=single]{siunitx}
\usepackage{siunitx} 
\usepackage{multirow} 
\dottedcontents{section}[3em]{\bfseries}{1.2em}{1pc}
\dottedcontents{subsection}[6em]{}{2.0em}{1pc}

\title{Edible microlasers for monitoring authenticity and quality of food and pharmaceuticals}

\author[1,2]{Abdur Rehman Anwar}
\author[1]{Maru\v{s}a Mur}
\author[3]{Georgia Michailidou}
\author[3]{Dimitrios N. Bikiaris}
\author[1,4,5]{Matja\v{z} Humar*}
\affil[1]{Department of Condensed Matter Physics, J. Stefan Institute, Jamova 39, SI-1000 Ljubljana, Slovenia}
\affil[2]{Jozef Stefan International Postgraduate School, Jamova 39, 1000, Ljubljana, Slovenia}
\affil[3]{Laboratory of Polymer Chemistry and Technology, Department of Chemistry, Aristotle University of Thessaloniki, Thessaloniki, Greece}
\affil[4]{CENN Nanocenter, Jamova 39, SI-1000 Ljubljana, Slovenia}
\affil[5]{Faculty of Mathematics and Physics, University of Ljubljana, Jadranska 19, SI-1000 Ljubljana, Slovenia}
\affil[ ]{*E-mail: matjaz.humar@ijs.si}

\date{}

\begin{document}
\maketitle

\textbf{Keywords:} microlasers, edible materials, food, pharmaceuticals, sensing, barcoding, anti-counterfeiting

\begin{abstract}
Traceability, security and freshness monitoring are crucial to the food and pharmaceutical industries. Currently, barcodes and sensors are almost exclusively located on product packaging. Making them edible and introducing them into edible products could significantly enhance their functions. Here, several types of microlasers made entirely out of edible substances were developed. It is striking that olive oil already contains enough chlorophyll to be used as a laser when dispersed in water as droplets. The edible lasers can be embedded directly into edible products and serve as barcodes and sensors. Due to their much narrower spectral lines compared to fluorescent or color-changing sensors, they are significantly more sensitive to various environmental factors. The edible lasers were employed to sense sugar concentration, pH, the presence of bacteria, and exposure to too-high temperatures. They can also encode tens of data bits, such as manufacturer's information and expiration date. The microlasers are entirely safe for consumption, do not change the appearance and taste of food considerably, and are environmentally friendly. The developed barcodes and sensors could also be applied to non-edible items, such as cosmetic and agricultural products, for environmental monitoring and biomedical applications.

\end{abstract}

\newpage

Counterfeiting and poor quality are rising concerns in the food and pharmaceutical industry, which is alarming since they are directly linked to public health concerns. There are various reliable conventional methods to evaluate food quality and authenticity, such as mass spectrometry, DNA barcoding, isotope and elemental fingerprints, and gas chromatography \cite{medina2019current,ye2023comprehensive}. However, these techniques require sophisticated instrumentation; they are time-consuming, expensive, inappropriate for consumer use, and usually require sample collection to be analyzed \cite{wadood2020recent,ye2023comprehensive}. The most widely used technique currently on the market to track products is to label them with barcodes, primarily by printing them on the packaging. However, the packaging can be easily removed or substituted. Therefore, such barcoding only partially ensures traceability and counterfeit prevention. It is also unsuitable for bulk products when they are still unpackaged.

To reduce or minimize counterfeiting, there is a demand for better tracking and authentication systems, such as those where food or pharmaceuticals themselves are labeled, not only their packaging. An edible barcode or sensor embedded directly into the product and remotely read by a small handheld reader would be ideal for these applications. A few solutions exist but are either only at a proof-of-concept stage, are only partially edible, or require complex readout methods. For example, erythrosine-doped corn starch microparticles randomly positioned in gelatin \cite{esidir2023food} and silk fibroin genetically encoded with fluorescent proteins \cite{leem2020edible,leem2022edible} were used as physically unclonable functions (PUFs). Another example are flexible and biocompatible PUFs, created by randomly positioning microdiamonds in silk fibroin film, which were read by measuring the Raman signal \cite{hu2021flexible}. Recently, water-soluble nanocomposite ink was used for labeling food products and was simultaneously used as a humidity indicator \cite{kim2024water}.

Together with traceability, the monitoring of food freshness is essential for consumers. Freshness sensors help reduce waste because the food is often good past the predetermined expiration date. Various sensors have been integrated into food packaging to address the issue, constituting so-called smart or intelligent packaging \cite{yousefi2019intelligent}. By actively monitoring multiple parameters inside the food packaging, such as pH, humidity, temperature, oxygen and carbon dioxide level, and microbial growth, these sensors provide information about the state of the food product \cite{weston2021food}. Primarily, these sensors are based on sensing colorimetric and electrochemical changes. However, they have sensitivity and lifetime issues and increase packaging waste \cite{dudnyk2018edible,siddiqui2022nanomaterials,rodrigues2021bio, morsy2016development}.

In addition to smart packaging, edible sensors were reported but mostly applied to biomedical research \cite{tao2012silk,sharova2021edible}. Examples include edible thermistors for measuring oral cavity temperature \cite{stojanovic2024temperature}, biodegradable and deformable temperature sensors for monitoring the temperature wirelessly \cite{salvatore2017biodegradable}, and electrochemical sensors for monitoring the glucose levels in sweat \cite{chen2022silk}, as well as other physiological parameters \cite{kim2017edible} and for detecting the defrosting event of frozen products \cite{ilic2022self}.

Microbarcodes based on microcavities and microlasers have gained a great deal of attention because of their potential in tracking, labeling, bio-detection, cell tagging, information security, and anti-counterfeiting \cite{anwar2023microcavity,shikha2017versatile,leng2015suspension}. Microlasers have a narrow emission spectrum and therefore have the potential to generate millions of unique tags \cite{dannenberg2021droplet,martino2019wavelength}. In recent years, they have been increasingly employed as barcodes \cite{anwar2023microcavity,humar2015intracellular,dannenberg2021droplet,martino2019wavelength,schubert2015lasing,hill2014advances} and microsensors \cite{he2011detecting,yu2021whispering, toropov2021review, wang2020microfluidic,schubert2020monitoring}, however, until now they have mainly been applied to biomedical applications \cite{fan2014potential, chen2019biological} and were not specifically made from edible materials. There are a few cases where either the cavity or the gain medium were made from bio-materials appropriate for consumption. For example, egg-white \cite{van2020egg} and bovine serum albumin \cite{van2019protein} were used as cavity materials in whispering gallery mode (WGM) lasers with a non-edible gain medium. Edible fluorescent dyes riboflavin (vitamin B$_2$) and chlorophyll were used as gain materials in a distributed-feedback (DFB) laser \cite{vannahme2013single} and in a WGM laser \cite{chen2016optofluidic,nizamoglu2013all}, where either the laser cavity material or the substrate on which it resided was not edible. 

A few lasers were also reported to be based on entirely edible materials, but they were not specifically designed to be edible or used in food. For example, a random laser based on marine materials was demonstrated \cite{lin2018all}, using chlorophyll extracted from marine diatoms as a gain medium and blue coral skeletons as the scattering medium. Also, a single-mode DFB laser made from silk, riboflavin and silver \cite{choi2015fully} was demonstrated. In another study, curcumin microspheres exhibited WGM resonances but not lasing \cite{venkatakrishnarao2016photonic}.

\begin{figure}[h]
	\centering
	\includegraphics[width=17cm]{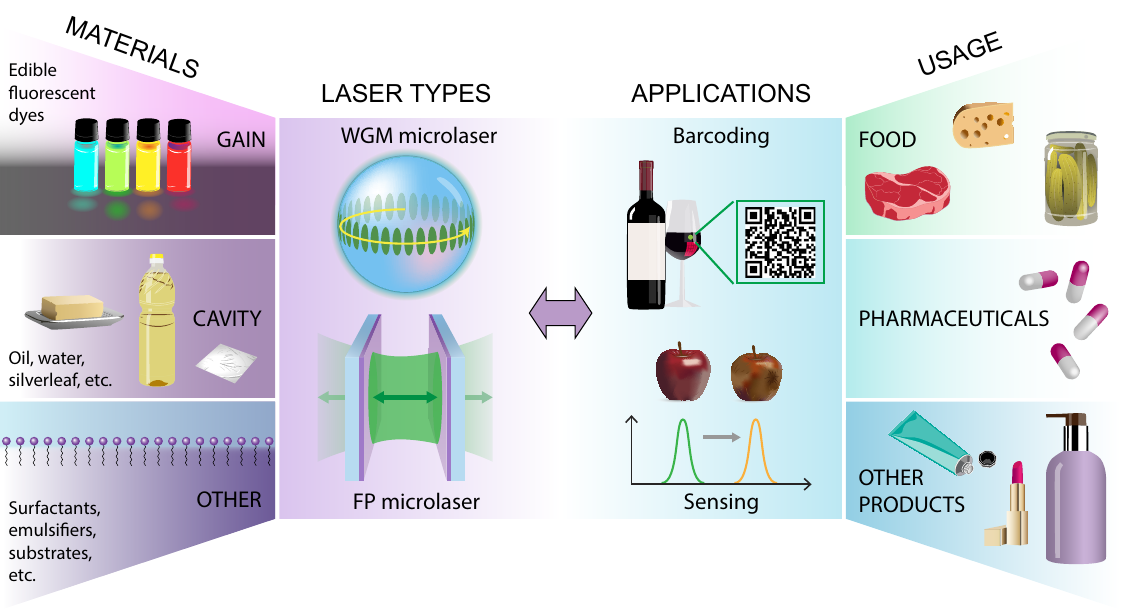}
	\caption{Development of completely edible microlasers in different configurations from materials to applications. Various edible laser gain materials, cavity materials and optional secondary materials were identified and employed to make two types of lasers: WGM and FP. These lasers were employed for barcoding and sensing, enhancing authentication, and reducing the health risks of various products.
}
	\label{fig:1}
\end{figure}

In this work we aimed to systematically search and test different dyes, cavity materials and geometries to make completely edible lasers and employ them for various applications related to food and pharmaceuticals for the first time (Fig.\ \ref{fig:1}). We demonstrated microlasers based on edible materials in several different configurations. We tested these edible lasers inside food and demonstrated their practical long-term use as barcodes, encoding useful information such as the expiration date. Additionally, we employed them as sensors to detect several parameters related to food quality and safety.

\subsection*{Basic concepts and constituents of edible microlasers}

Lasers consist of three primary building blocks: a gain medium, an optical microcavity, and an energy source. In our case, the gain medium, a fluorescent dye, provides optical gain through stimulated emission. Based on their light confinement mechanism, microcavities can be of different types. We demonstrated two types, whispering gallery mode (WGM) and Fabry–Pérot (FP) microlasers. The microlasers are pumped with an external light source, such as a pulsed laser. When the total optical gain in the cavity exceeds the total optical losses, the system reaches the lasing threshold, resulting in laser emission. Here, we made both the gain medium and the cavity exclusively out of edible substances, either those naturally occurring in food or those specifically approved to be used as food additives, without any chemical alterations. 

Many food colorants are available, and among them, we searched for dyes with sufficiently strong fluorescence to be used as the gain material. We achieved lasing using four edible dyes: chlorophylls, riboflavin, riboflavin sodium phosphate and bixin (Table 1). The dyes with a quantum yield below $\sim0.2$ were inappropriate for lasing. In some cases, even though a dye had a high enough quantum yield, the limitation was its solubility in solvents commonly contained within food (e.g. water and oils). We did not consider pure ethanol as a solvent since it would not be practical to use it due to fast evaporation or diffusion into the surrounding food. Further, we did not consider dyes that are safe to ingest but are not approved for food use. For example, fluorescein is an excellent gain material and is non-toxic, but it is only approved for oral use for medical purposes. Also, fluorescent proteins (e.g., GFP) are present in nature and have been used for various microlasers \cite{gather2011single,gather2014bio}, but are not allowed for human consumption as an additive and, therefore, have not been used in this study.

\begin{table}
\small
\setlength{\tabcolsep}{5pt}
\begin{tabular}{ccccp{3.5cm}} 
 \toprule
 \textbf{Dye} & \textbf{Pumping/lasing}  & \textbf{QY} &
 \textbf{Solvent} & \textbf{Laser type}\\ 
 \midrule
 \multirow{3}{*}{Chlorophyll-A} & \multirow{3}{*}{440 or 525/\SI{680}{\nano m}}& \multirow{3}{*}{0.3\cite{leupold1990excited}} &  \multirow{2}{*}{vegetable oil} & WGM (oil droplets in water) \\
 && & &  FP\\
 \cmidrule{4-5}
 & & &cooked butter & WGM (solid beads in water)\\
 \midrule
\multirow{2}{*}{Chlorophylls in olive oil} & \multirow{2}{*}{440 or 525/\SI{680}{\nano m}}& & & WGM (oil droplets in water) \\
 \cmidrule{5-5}
 (naturally present) & & & & FP \\
 \midrule
 Non-purified chlorophyll & 440 or 525/\SI{680}{\nano m} & & vegetable oil& WGM (oil droplets in water)\\
 \midrule
 Riboflavin (vitamin B2) &  450/\SI{550}{\nano m}& 0.27\cite{drossler2002ph} & water-glycerol  & WGM (droplets on a superhydrophobic surface)\\
 \cmidrule{1-5}
 \multirow{2}{*}{Riboflavin sodium phosphate}  & \multirow{2}{*}{450/\SI{550}{\nano m}}& \multirow{2}{*}{0.23\cite{holzer2005photo}} & water  & FP \\ 
 \cmidrule{4-5}
  & & &  water-glycerol & WGM (droplets on a superhydrophobic surface) \\ 
 \midrule
 Bixin & 450/\SI{530}{\nano m} &  &vegetable oil & WGM (oil droplets in water)\\
 \bottomrule
\end{tabular}
\caption{List of demonstrated edible lasers including dye used,  pumping and central lasing wavelengths, quantum yield, solvent and laser type.}
\label{table1}
\end{table}

The choice of resonator material depends on the specific microlaser configuration and function. Typically, the materials should be transparent and, in some configurations, need to have a high refractive index or be reflective if used as a mirror. We used various oils, butter, agar, gelatin, chitosan and thin silver leaves, to produce cavities. Additional materials were used for non-optical components, such as for mechanical support, encapsulation or to stabilize droplet emulsions.

In general, none of the substances used here were chemically modified in any way. They were used in reasonable quantities and forms, commonly present in food and pharmaceuticals. Therefore, products' visual appearance, taste, or nutritional value were not changed considerably and remained environmentally friendly.

\subsection*{Edible whispering gallery mode lasers}

\begin{figure}[!tb]
	\centering
	\includegraphics[width=15cm]{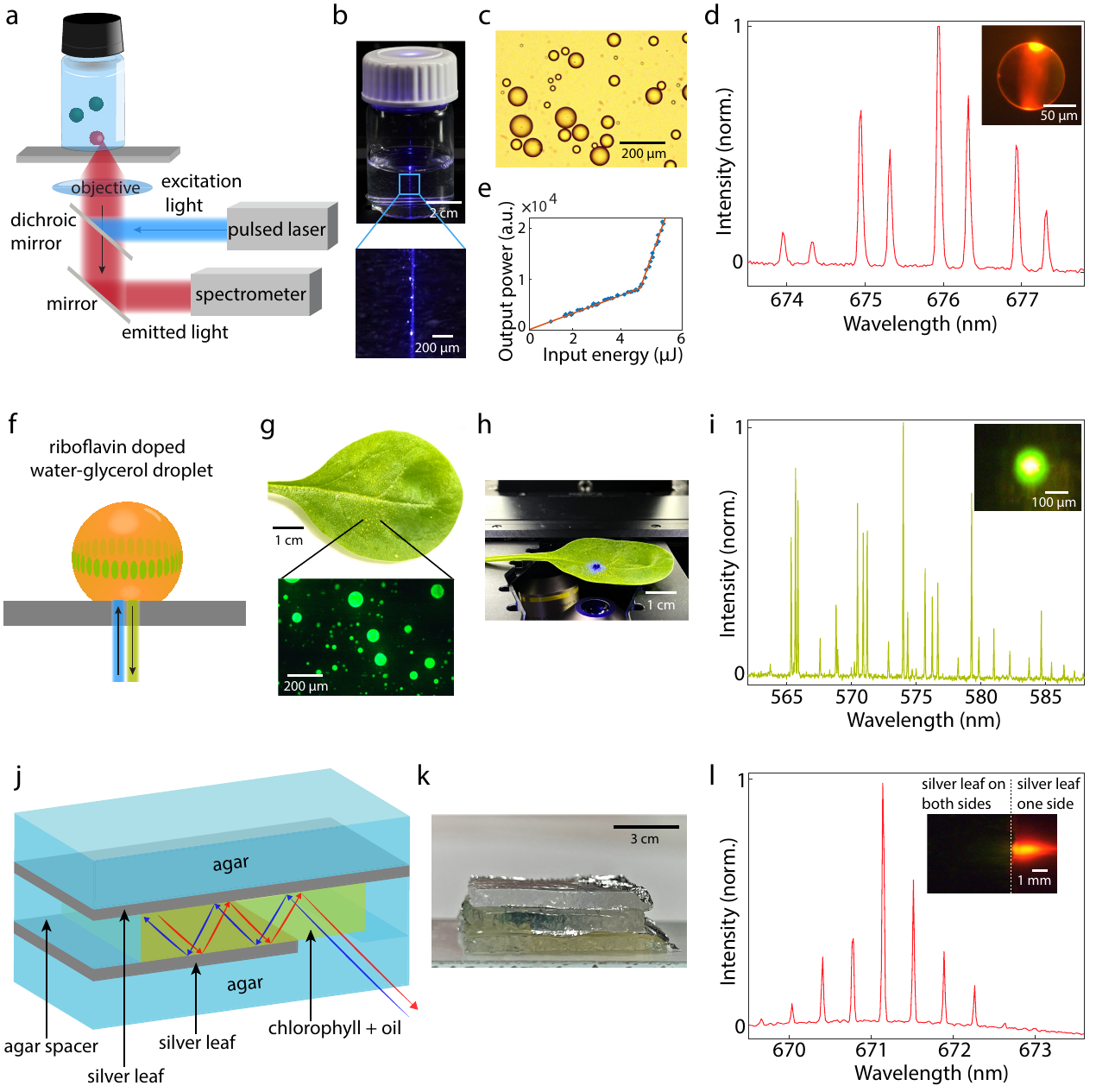}
	\caption{ a) Schematics of the experimental setup, which includes a glass bottle filled with chlorophyll-doped oil droplets dispersed in water and an optical system for the excitation and signal collection. b) Chlorophyll-doped oil droplet WGM lasers under the excitation of a blue pulsed laser. c) Bright-field image of polydispersed chlorophyll-doped oil droplets. d) Emission spectrum from a single chlorophyll-doped oil droplet with a diameter of \SI{\sim100}{\micro m} shown in the inset. e) The output power of a droplet laser shows typical threshold behavior as the input pulse energy is increased. f) Scheme of an edible WGM laser as a dye-doped droplet on a superhydrophobic surface. g) Riboflavin-doped water-glycerol solution was sprayed onto a spinach leaf and the fluorescence of the resulting droplets was imaged.  h) The spinach leaf was positioned on the microscope and excited by a blue pulsed laser to achieve lasing. i) Laser emission spectrum from a single droplet shown in the inset. j) Schematics of an edible FP cavity laser. The cavity is filled with chlorophyll-doped oil and illuminated by a pulsed laser. The excitation/emission light enters/exits at the position where the silver leaf is only on one side of the cavity. k) Side view of an edible FP cavity laser. l) Laser emission spectrum from an edible FP cavity shown in the inset.
}
	\label{fig:2}
\end{figure}

The simplest lasers to make are WGM-based microlasers. They are composed of dye-doped droplets or solid spheres in an environment with a lower refractive index. Such microcavities confine light via multiple total internal reflections on the smooth surface \cite{vahala2003optical,vollmer2008whispering}. WGMs typically have very high Q-factors and, therefore, low lasing thresholds, and they do not require the dyes to have a very high quantum yield. For this reason, we used WGMs to test possible candidate dyes. Specifically, the oil-soluble dyes were tested in sunflower oil droplets dispersed in water (Fig.\ \ref{fig:2}a). The water-soluble dyes were tested in water droplets on a superhydrophobic surface. 

We achieved lasing with \SI{2}{\milli M} chlorophyll-A or \SI{4}{\milli M} bixin dissolved in oil. Edible emulsifier polysorbate was added to the water phase to stabilize the droplet dispersion. Polydispersed droplets with a size range \SIrange{20}{120}{\micro m} were produced by vigorously shaking the bottle containing both phases (Fig.\ \ref{fig:2}b and c).

The droplets were pumped with a tunable nanosecond pulsed laser (Fig.\ \ref{fig:2}b and Supplementary Fig. 1a) at \SI{440}{nm} or \SI{450}{nm}, for chlorophyll and bixin, respectively. The light was sent to a camera or a spectrometer. Sharp spectral lines corresponding to WGM modes with TE and TM polarizations were observed in the spectrum of the emitted light (Fig.\ \ref{fig:2}d). The laser's output power as a function of input energy displays the typical threshold behavior, which indicates lasing (Fig.\ \ref{fig:2}e). The chlorophyll- and bixin-doped droplets had a threshold energy of \SI{4.7}{\micro J} and \SI{6.8}{\micro J}, respectively. In both cases, the minimum droplet size required to achieve lasing was approximately \SI{35}{\micro m}. This limitation originates in the radiation leakage that primarily depends on the contrast between the internal and external refractive index and on the size of the cavity. Instead of pure chlorophyll-A, a non-purified mixture of chlorophyll extracted from spinach by ethanol also emitted laser light from oil droplets in water. Further, even pure olive oil already naturally contains enough chlorophyll to enable lasing as droplets. However, when using non-purified chlorophyll or olive oil, the laser threshold is larger, typically by a factor of 3. Notably, WGM peaks in the spectrum can also be observed below the lasing threshold (Supplementary Fig. 2a and b). This means that WGM microcavities can also be used with dyes that do not provide enough gain for lasing. In this case, a continuous wave laser or an LED can be used for the excitation.

Water droplets on a superhydrophobic surface were employed to test the water-soluble dyes (Fig.\ \ref{fig:2}f). Many plant leaves are superhydrophobic \cite{lee2006lotus, neinhuis1997characterization}. We used spinach leaves as a superhydrophobic surface to make a completely edible laser. Riboflavin and riboflavin sodium phosphate were dissolved in a water-glycerol mixture at \SI{3}{\milli M} and \SI{5}{\milli M} concentration, respectively. Glycerol was added to reduce the evaporation rate. The solution was sprayed onto a spinach leaf using a spray bottle (Fig.\ \ref{fig:2}g), forming polydisperse microdroplets. The leaf was flipped upside down and placed onto an inverted microscope. The microdroplets emitted green fluorescent light upon excitation with a blue LED (Fig.\ \ref{fig:2}g). One droplet was pumped with a nanosecond pulsed laser at \SI{450}{nm} (Fig.\ \ref{fig:2}h). Above the lasing threshold, narrow irregular peaks were observed in the emitted light (Fig.\ \ref{fig:2}i). While this is an easy way to achieve lasing, it is not practical for real-life applications since the water droplets evaporate, can come into contact with food and can be easily displaced from the superhydrophobic surface.

\subsection*{Edible Fabry–Pérot lasers}

The second type of an edible laser was based on a Fabry–Pérot resonator, which is a linear cavity consisting of two mirrors with a gain medium between them \cite{kavokin2017microcavities}. The most reflective edible materials are silver, gold and aluminum. They are usually used as decorations for some foods and drinks in the form of extremely thin leaves. We developed an edible laser by attaching the silver leaves to agar layers that provided structural support (Fig.\ \ref{fig:2}j and k). An additional agar layer was placed between the two mirrors as a spacer, but only at the edges. The space in between the mirrors was filled with \SI{2}{\m M} chlorophyll solution in oil or with \SI{5}{\m M} riboflavin sodium phosphate solution in water. One mirror was made intentionally shorter to enable the excitation light to enter the cavity and the resulting laser light to exit the cavity. Upon pumping the cavity with a pulsed laser, sharp, equally spaced peaks appeared in the emission spectrum, indicative of lasing within the FP cavity (Fig.\ \ref{fig:2}l).

\subsection*{Edible microlaser barcodes}

\begin{figure}[!b]
	\centering
	\includegraphics[width=15cm]{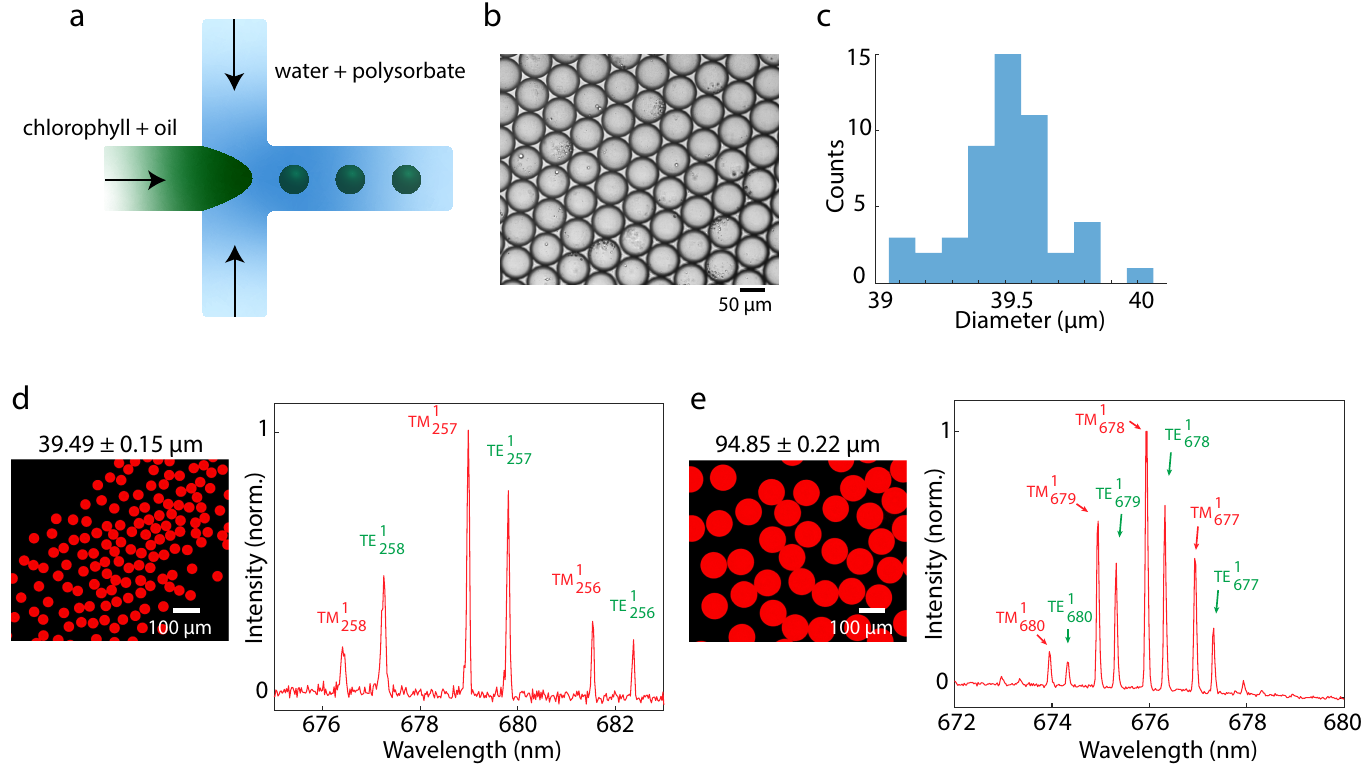}
	\caption{a) Experimental setup for the generation of monodispersed droplets. b) Bright-field image of the monodispersed chlorophyll-doped oil droplets. c) Size distribution of the droplets of one sample. d) Fluorescence image of monodisperse chlorophyll-doped oil droplets (mean diameter \SI{39.49 \pm 0.15}{\micro m}) and lasing spectrum of a single droplet. e) Fluorescence image of monodisperse chlorophyll-doped oil droplets (mean diameter \SI{94.85 \pm 0.22}{\micro m}) and lasing spectrum of a single droplet.
}
	\label{fig:3}
\end{figure}

We employed the WGM droplet lasers for barcoding. The size of a WGM microcavity can be determined from its emission spectrum with a nanometer accuracy \cite{richter2020optical, kavcic2022deep}, and can be used as a unique barcode. Most previous studies used polydispersed microlasers to generate random barcodes \cite{martino2019wavelength, humar2015intracellular,fikouras2018non}. Here, we instead produced lasers with precisely determined sizes to encode information \cite{richter2020optical}, like with regular barcodes. Using a microfluidic droplet generator chip we generated highly monodispersed chlorophyll-doped oil droplets in water (Fig.\ \ref{fig:3}a and b). A typical size distribution of the droplets produced in this way is shown in Fig.\ \ref{fig:3}c. The water phase contained a small amount of polysorbate as an edible emulsifier. By changing the pressures of the water and oil phases, the droplet size could be precisely controlled. Different sizes of droplets were collected in separate vials to be later used as barcodes (Fig.\ \ref{fig:3}d and e).

Upon excitation with a pulsed laser, the typical WGM lasing spectrum dependent on the droplet size could be observed (Fig.\ \ref{fig:3}d and e). Pairs of peaks were visible, corresponding to TE and TM polarizations. The mode wavelengths $\lambda_l$ can be approximated by $2\pi n R =l \lambda_l$, where $R$ is the cavity radius, $n$ is the effective refractive index inside the cavity and $l$ is the mode number. From this equation, the approximate size of the cavity can be determined. To calculate the size more accurately, the peak positions were fitted to the exact WGM solutions \cite{eversole1993high}. The fitting parameters were the size of the droplet and the external refractive index, while the internal refractive index was known in advance (1.445, measured with a refractometer). A change in the refractive index of the surrounding medium results in a shift of the modes; however, the fitted diameter, which encodes the barcode, will remain unchanged. Therefore these barcodes are robust to the change of the environment. The diameter was fitted with an absolute error of about \SI{70}{\nano m}. By measuring about 50 droplets, their size coefficient of variation (CV, defined as the ratio of standard deviation to the mean of the droplet diameter) was found to be typically \SIrange{0.2}{0.4}{\percent}. 

\begin{figure}[!b]
	\centering
	\includegraphics[width=15cm]{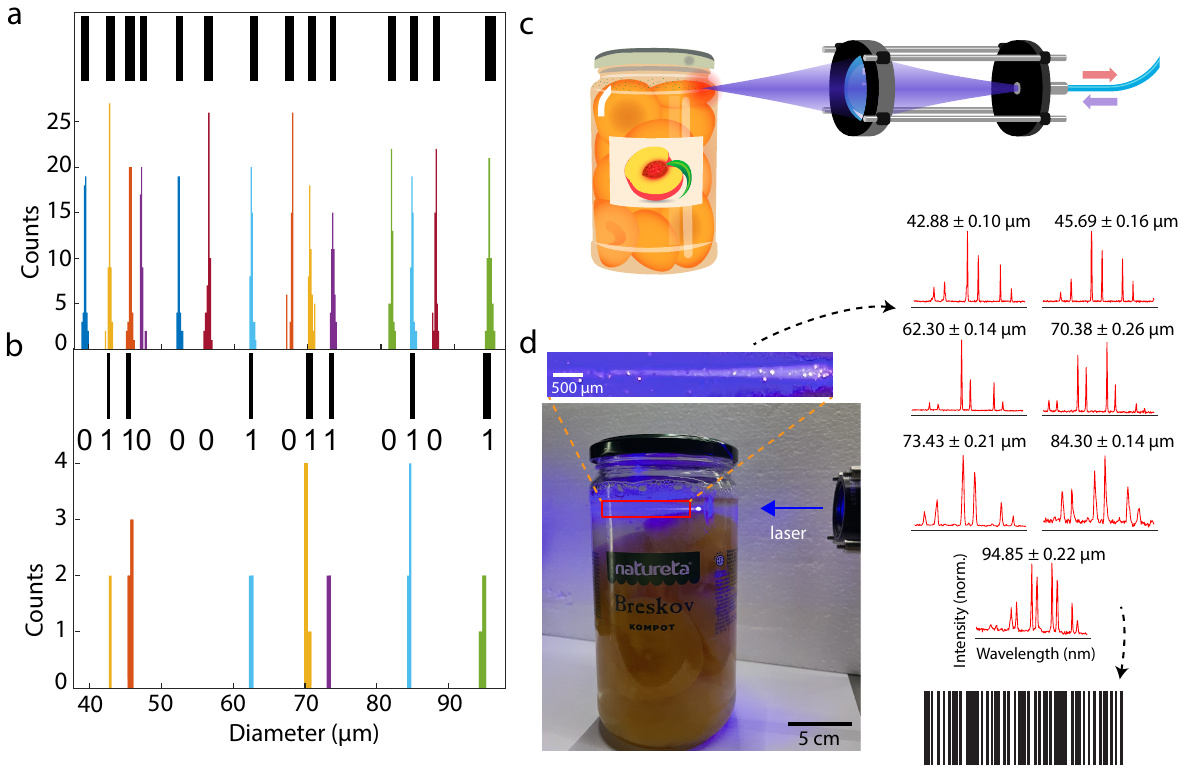}
	\caption{ a) Size distribution of chlorophyll-doped oil droplets of 14 different diameters produced by a droplet generator chip. b) Distribution of droplet sizes detected in the jar after the date of "Stop Food Waste Day" was encoded by adding selected droplet sizes into the sample. All sizes were reliably detected. c) Schematics of the experimental setup for reading the barcodes, which includes an optical fiber for excitation and collection of the emission signals. d) Reading of an edible barcode based on chlorophyll-doped oil droplets in a jar of compote. The food product is scanned by a nanosecond pulsed laser at \SI{440}{nm}. Droplet sizes are calculated from the acquired spectra and the corresponding barcode is reconstructed. 
}
	\label{fig:4}
\end{figure}

To generate a barcode we produced 14 different sets of monodispersed droplets, with non-overlapping sizes in the size range \SIrange{39}{95}{\micro m} (Fig.\ \ref{fig:4}a). A total of 14 bits of information can be encoded, as each size can be present in the food sample (1) or not (0). In this way, $2^{14} = 16.384$ unique strings of binary digits can be produced. Information such as manufacturing or expiry date, country of origin, and manufacturer identification could be encoded into food and pharmaceutical products.
 
We encoded a specific date into a peach compote in a glass jar to demonstrate the barcoding technique. We chose the first international "Stop Food Waste Day," commemorated on April 26, 2017. January 1, 2000 was set to be the starting day for the date encoding scheme. This date was encoded as 00000000000001. With each unique bit-string representing one day, approximately 44 years can be covered. The first "Stop Food Waste Day" was commemorated 6325 days after the starting date of our encoding scheme. When converted into a binary string of 14 digits, 6325 equals 0110001010101 (Fig.\ \ref{fig:4}b).

In our experimental demonstration, each droplet size corresponded to a position in the bit string. Droplets, for which the bit value corresponded to 1, were added to the glass jar. To read the barcode, the glass jar was scanned with a nanosecond pulsed laser at \SI{440}{nm} and the emission spectra were recorded (Fig.\ \ref{fig:4}c). When a droplet entered the illumination path, sharp lines corresponding to WGM lasing appeared in the collected spectrum (Fig.\ \ref{fig:4}d). The size was calculated for each spectrum and plotted into a histogram (Fig.\ \ref{fig:4}b). All the sizes we inserted into the glass jar were successfully detected, with an average of 5 lasers of each size being detected during a few-second measurement. While here the lasers were very sparse and we mostly detected a signal from just one at a time, it could be possible to read more of them simultaneously and use unmixing algorithms to extract their individual sizes from a single spectrum, as shown before \cite{kavcic2022deep}, thus increasing the readout speed. 

The amount of encoded information can be further increased by increasing the interval of droplet sizes and decreasing the separation of sizes. The smallest size of a droplet laser is limited by the radiative leakage, which is ultimately limited by the refractive index. By using higher refractive index oils such as cinnamon oil ($n=1.55$), the minimum size to achieve lasing can be decreased down to \SI{30}{\micro m}. An additional benefit of cinnamon oil is its density, which is higher than water, so that the droplets will remain suspended inside the aqueous medium rather than floating at the surface. Further, cinnamon oil can be mixed with regular oils in the right ratio to match the density of water. The largest droplet size is, in principle, not limited, but above \SI{\sim100}{\micro m}, higher radial WGM modes start to appear, which makes the fitting of the sizes more complicated. In the size range \SIrange{39}{95}{\micro m} and a CV of \SI{1.9}{\micro m}, about 30 distinct sizes can be fitted without much overlap (Supplementary Fig. 3b), enabling the encoding of 30 bits of information.

Finally, we successfully detected lasing spectra from droplets mixed into various other grocery items, including a pickle jar and a juice bottle (Supplementary Fig. 4). The droplet lasers in water were very stable. The barcode could be read for more than 18 months after it was produced by storing it at room temperature on a shelf in a glass container without any special light protection.

\subsection*{Sugar concentration measurement}

Laser microcavities are excellent for sensing applications because of their extreme sensitivity to changes in cavity properties, including their size, shape, refractive index and the surrounding medium. The microcavities can be functionalized using responsive materials to sense various parameters \cite{kavcic2022deep}. Here, we employed edible WGM and FP lasers for sensing applications, including measuring sugar level, pH, temperature, and presence of bacteria. These sensors can be placed directly inside the food or food packaging for real-time monitoring of food freshness and quality.

Refractive index measurement is typically used to determine sugar concentration in the production of various foods, such as wine, beer, maple syrup, and honey. A sample must be taken and put into a refractometer to measure the refractive index. If such a sensor was present inside the fluid itself, it would be possible to continuously monitor sugar concentration, including in a sealed container, such as in a bottle where fermentation occurs. WGMs are perfectly suitable for this purpose since the absolute refractive index of the surrounding medium can be determined \cite{zheng2018sensing,merabet2023glycosuria}. Both solid or liquid WGM lasers can be used for this purpose. We employed chlorophyll-doped sunflower oil droplets to demonstrate sugar concentration measuring with edible WGM microlasers. We measured the external refractive index for several different glucose concentrations. Fitting of the droplet size and the external refractive index was performed for the spectra of each droplet and the external refractive index was averaged over five droplets from each sample (Fig.\ \ref{fig:5}a). The average standard deviation of the refractive index measurement was 0.0003 RIU, which enabled the measurement of the glucose concentration in the solution with an error of 0.2 percentage points. This is comparable to standard commercial refractometers.

\subsection*{pH sensors}

\begin{figure}[!tb]
	\centering
	\includegraphics[width=15cm]{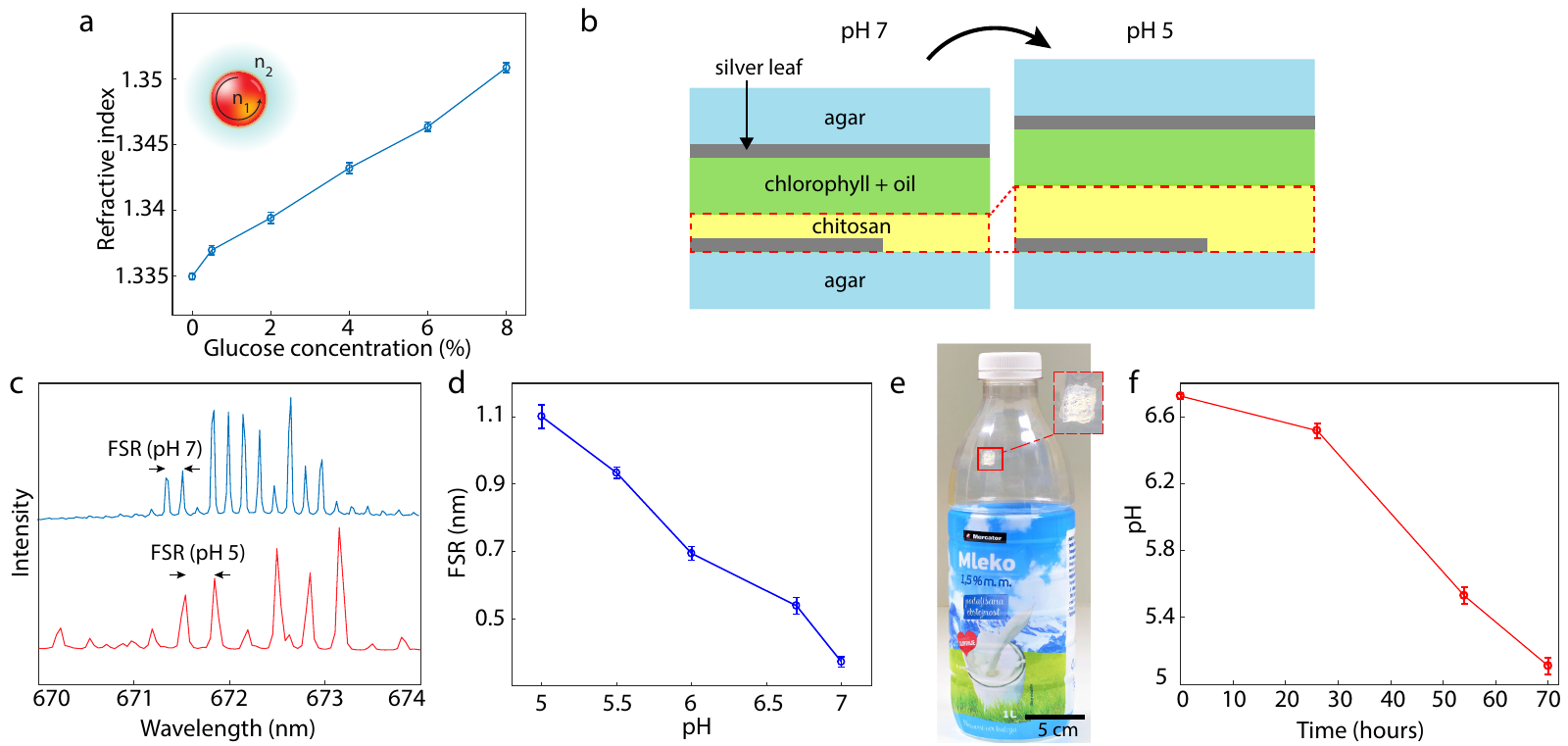}
	\caption{a) The measured refractive index of different concentrations of glucose in water. The index is measured via the lasing spectrum of the droplet lasers. b) Schematic diagram of the edible FP cavity-based pH sensor at two different pH. The chitosan layer swells as the pH is reduced, changing the emission of the laser. c) Emission spectra from the FP cavity in (b) pumped with nanosecond pulsed laser at \SI{440}{nm} at two different pH. d) The change of the FSR with respect to the pH change. e) The edible FP laser pH sensor inside the milk bottle measures the pH of the milk. f) The change of the milk's pH over two days when left at room temperature indicates the spoilage of the milk.
}
	\label{fig:5}
\end{figure}

We demonstrated a completely edible pH sensor that can be used to measure the pH of food in real time. The sensor is based on the edible FP cavity described previously. In the case of an FP cavity, the free spectral range can be calculated as $\mathrm{FSR} = (2 n L)^{-1}$, where $n$ is the refractive index within the cavity and $L$ is the cavity thickness. From the measured value of the FSR (Fig.\ \ref{fig:2}l) we were able to calculate the thickness of the optical cavity of a \SI{300}{\micro m} thick cavity with an error of \SI{600}{nm}. To make the FP microlaser pH sensitive, an additional chitosan film was introduced in between the edible mirrors (Fig.\ \ref{fig:5}b). Chitosan is a pH-sensitive material that swells as the pH decreases (Supplementary Fig. 5) \cite{bhattarai2010chitosan,aider2010chitosan}. In our sensor, the surrounding liquid diffuses through the agar layer to the chitosan layer. If a change in pH occurs, the thickness and refractive index of the chitosan film change. This varies the optical path length in the cavity, which results in a change of the free spectral range of the laser emission (Fig.\ \ref{fig:5}c and d).

To test the pH sensitivity of the edible sensor, it was initially immersed in a pH 7.2 solution for approximately 1 hour. Later, the laser was immersed in different pH solutions down to pH 5. The FSR decreased approximately linearly with increasing pH (Fig.\ \ref{fig:5}d), indicating the decrease in the optical cavity length as expected. The error in measuring pH was 0.05 pH units, which was estimated by repeating the measurements.

To demonstrate a practical application of the edible pH sensor, we tested its performance by measuring the pH of milk as it got spoiled when kept at room temperature for several days. The sensor was fixed inside a milk bottle by using an edible sealant (Fig.\ \ref{fig:5}e).  
The emission spectrum of the FP microlaser was collected multiple times over 70 hours. By taking into account the previously measured dependence of FSR to pH (Fig.\ \ref{fig:5}d), the change of pH in time was determined (Fig.\ \ref{fig:5}f). These results agree well with the usual drop in pH when milk gets spoiled \cite{poghossian2019rapid, pesta2007effects}.

\subsection*{Detection of microorganism growth}

\begin{figure}[!b]
	\centering
	\includegraphics[width=11cm]{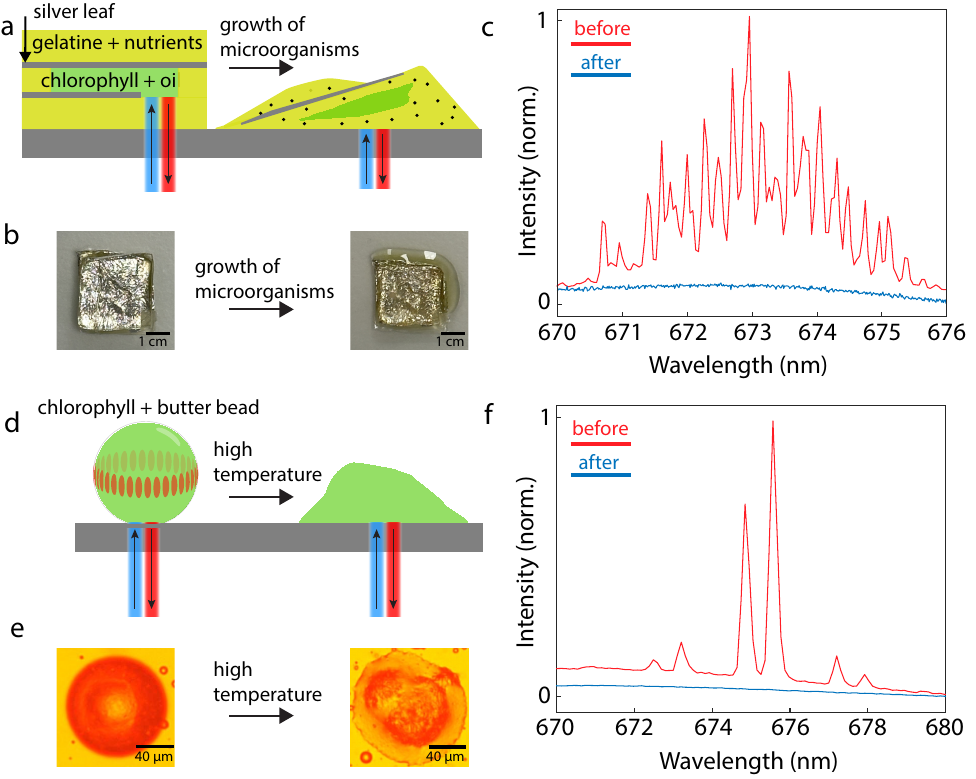}
	\caption{a) Schematic diagram of the microlaser-based microorganism sensor before and after the growth of microorganisms. b) Images of the microorganism sensor before (day 0) and after (day 5) the growth of bacteria (top view). c) Emission spectra from the microorganism sensor before (day 0) and after (day 5) the growth of bacteria. d) Schematic diagram of the temperature sensor at low and high temperatures. e) Bright-field images of the temperature sensor based on chlorophyll-doped butter bead at low and high temperatures. f) Emission spectra from temperature sensor at room temperature and after it has been exposed to a high temperature (\SI{35}{\degreeCelsius}).
}
	\label{fig:6}
\end{figure}

We developed an FP cavity-based sensor that can be placed inside the food packaging, such as within the meat package, to detect the growth of microorganisms. The FP laser was made as described before (Fig.\ \ref{fig:2}j) from edible silver leaf, and non-purified chlorophyll-doped oil solution, except that the agar layers were replaced by nutrient-enriched gelatin (Fig.\ \ref{fig:6}a). Upon pumping with the nanosecond pulsed laser at \SI{525}{\nano m}, peaks appeared in the emission spectrum, indicating the FP-cavity lasing. During its preparation, the sensor was contaminated with microorganisms from the air, leading to microorganism growth in gelatin in the following days. Some bacteria produce the enzyme gelatinase, which digested the gelatin layers within several days, resulting in liquefaction of the medium. This caused the laser cavity to fall apart and not support lasing anymore (Fig.\ \ref{fig:6}b and c). The laser signal's disappearance indicated significant microbial growth. The bacterial growth is highly dependent on the temperature. Therefore, such a sensor could also indicate if the food was kept at an elevated temperature for a longer time. In some cases mold also started growing on the edible lasers, which can block the light in the cavity and indicate food spoiling.

\subsection*{Microlaser temperature sensors}

We developed an edible temperature sensor using a WGM laser. The sensor changes irreversibly when exposed to a specific temperature (Fig.\ \ref{fig:6}d). This sensor can be used to determine whether the food has been exposed to high temperatures or has been unfrozen. It can also be used to indicate the opposite if the food has been adequately heat-treated, which is crucial for eliminating harmful pathogens. Our sensor was made from chlorophyll-doped butter beads, which exhibited WGM lasing modes upon pumping with a pulsed laser (Fig.\ \ref{fig:6}e and f). When the temperature increased beyond butter's melting point at approximately \SI{35}{\degreeCelsius}, the beads deformed and the lasing ceased. Once the cavity is deformed, it cannot be reversed. Therefore, it is a good indication that the temperature was above the melting point at some point during the transport or storage of food or pharmaceutical products.

The cavity materials can be selected based on the melting temperature according to the desired application. Unsaturated fats, such as vegetable oils, have low melting temperatures, usually below freezing. In contrast, predominately saturated fats, such as butter, lard, margarine and palm oil, have melting temperatures typically above room temperature. The fatty acids that are the main constituents of fats have a very broad range of melting temperatures from \SI{-45}{\degreeCelsius} to \SI{80}{\degreeCelsius} \cite{knothe2009comprehensive}. Any melting temperature in this range can be selected by preparing the appropriate mixture. Alternatively, edible waxes can be used instead of fats.

\subsection*{Conclusion}

In this work, we demonstrated several edible lasers and their applications to enhance food and pharmaceutical security. It is the first time that edible laser dyes and microcavities were systematically studied. We identified four dyes and two laser cavity types that enable lasing. We demonstrated barcoding and sensing of sugar, pH, bacteria and temperature.

We searched through all the dyes allowed as food additives and tried about 15 dyes that we thought had strong fluorescence (Supplementary Table 1). Despite this, there are definitely more edible dyes that could support lasing. Scopoletin and quinine have a quantum yield of 0.56 \cite{pham2020fluorescence} and 0.58 \cite{nawara2019goodbye}, respectively, which should be more than enough for lasing. However, they have a low solubility in common solvents, so we could not achieve lasing in this case. There are many naturally occurring dyes for which the quantum yield is unknown, so the only way is to test them experimentally. Further, fluorescent proteins are optimized for a very high quantum yield and photostability and have been shown as perfectly suitable as a laser gain material \cite{gather2014bio, gather2011single, jonavs2014vitro, chen2014optofluidic}. They are currently not allowed for food or pharmaceutical use. However, they are believed to pose minimal toxicity and allergenicity when ingested \cite{richards2003safety, leem2020edible, jang2016single}.

WGM microcavities can be operated below the lasing threshold, that is in the spontaneous emission regime. In this case, they still show typical spectral lines and, therefore, can be used for barcoding and sensing, although at the expense of a lower signal. In this case, any fluorescent dye can be used, even if it has a lower quantum yield, although a very low quantum yield results in a very low signal. This opens the possibility of using many edible dyes. Examples are curcumin, nicotinamide adenine dinucleotide (NADH), vitamin E and A$_1$ and porphyrins. For example, curcumin microspheres were shown to support WGMs \cite{venkatakrishnarao2016photonic}.

Apart from WGM and FP cavities, there are also other possibilities such as distributed feedback (DFB) lasers \cite{vannahme2013single} and random lasers \cite{lin2018all}. DFB lasers have been employed before for sensing \cite{lu2008plastic,heydari2014label} and barcoding \cite{karl2018flexible}, so they could be good candidates for the applications demonstrated here.

Here, strictly materials allowed for use in food were used. However, there is a huge selection of non-toxic, biocompatible and biodegradable materials, which are used for medical purposes either for ingestion, injection or implantation, which could be employed for many different functional biolasers. Two examples are fluorescent dyes fluorescein and indocyanine green (ICG), which work very efficiently as laser dyes and are allowed for medical use but not as an additive in food. Therefore, some of the concepts developed within this work could also be applied to biomedical applications.

A critical application of the edible lasers demonstrated here is barcoding to label products and for anti-counterfeiting. One disadvantage of oil droplet-based barcodes is that they can only be used in water-based products but not in oil-based and solid products. However, the droplets generated in microfluidics could also be turned into solid spheres. There are two possible ways of making solid spheres from droplets. Firstly, a high melting temperature fat or wax could be used instead of the oil phase in microfluidics, but the droplets should be made at high temperatures. Upon lowering the temperature, the droplets would solidify \cite{lee2015shape}. Secondly, the droplets could be produced from a material dissolved in a solvent, which would afterward diffuse out leaving solid microbeads \cite{kim2014droplet}.

Parallelized microfluidic setups can generate monodispersed droplets at a fairly high rate \cite{gelin2020microfluidic, nisisako2008microfluidic}, which could be already enough to be practically useful in the production of high-value products such as pharmaceuticals. For large-scale industrial production, however, chemical methods to produce highly monodispersed beads \cite{ghimire2021renaissance, qiu2019synthesis} could be more appropriate. One example is silicon dioxide (silica) microparticles, which can be manufactured inexpensively in large quantities. Silicon dioxide is allowed as a food additive and is used as an anti-caking agent \cite{efsa2018re}. For example, \SI{50}{\micro m} silica microbeads can be produced with a CV of \SI{0.5}{\micro m}, which is worse than the CV of droplets, but still enough to encode tens of bits of information. A dye can be coated onto silica beads to produce WGM emission \cite{humar2017whispering} or possibly lasing.

Alternatively, instead of manufacturing lasers with precisely defined sizes to encode some predefined information, the barcode could be random. This is usually the case for laser barcodes used for cell barcoding \cite{humar2015intracellular,martino2019wavelength,schubert2015lasing}. In this case, each product would contain several randomly sized lasers and, therefore have a unique code, which would be read on the production line and stored in a database for later use. Since the size of WGM lasers can be measured down to \SI{1}{nm} precision \cite{humar2015intracellular}, but they can be manufactured only with a precision of \SI{100}{nm}, such barcodes would be unclonable. Such physically unclonable functions (PUF) would provide the ultimate anti-counterfeiting measure for high-value products.

Here, we demonstrated a few proof-of-concept sensors, however, the microlasers enable extremely precise and sensitive detection of any physical, chemical or biological process that changes the refractive index or physical dimension of the laser or other optical properties such as scattering, absorption and fluorescence. Therefore, we believe there are opportunities for detecting various other parameters related to food safety and freshness by searching for natural or developing different responsive edible materials.

In conclusion, we demonstrated several edible lasers and their applications. Since this is the first such study, there are many open possibilities for the development of various edible lasers and their applications, which could ultimately find their way to everyday use.

\subsection*{Materials and Methods}

Chlorophyll-A (Sigma Aldrich) was dissolved in ethanol and then added to sunflower oil (n=1.445). The ethanol was left to evaporate to yield a final dye concentration of \SI{2}{\milli M}. Deionized water was used as a continuous phase by dissolving \SI{1}{\percent} of polyoxyethylene (20) sorbitan monolaurate (Sigma Aldrich) as a surfactant to stabilize the droplets. Subsequently, \SI{1}{\percent} of the chlorophyll-doped oil was added into the water phase and vigorously shaken to produce droplets. Butter beads were produced in the same way just by working at a higher temperature, so the butter was liquid during the whole procedure.

Alternatively, fresh spinach leaves were used to extract the chlorophyll. They were washed with hot water and dried at \SI{50}{\degreeCelsius} for \SIrange{3}{4}{\hour}. The leaves were crushed in the presence of acetone with the help of a pestle and mortar. The solution was filtered and centrifuged for \SI{15}{\minute} at 2300 $g$. Then, hexane was added to the centrifuged solution with the ratio 7(hexane):3(centrifuged solution). The acetone and hexane formed separated layers after centrifugation (for 5 minutes at 2300 $g$). The hexane containing chlorophyll was collected and placed on the hot plate overnight at \SI{60}{\degreeCelsius}, for the evaporation of hexane. The resulting chlorophyll was dissolved into ethanol for further use.

The FP cavity was assembled from agar and thin silver leaves purchased at a baking supply store. The agar layers were prepared by dissolving \SI{1}{\g} of agar (Sigma Aldrich) in \SI{100}{\g} of water. The solution was sterilized by autoclaving at \SI{121}{\degreeCelsius} for \SI{20}{\min}, cooled down to \SI{65}{\degreeCelsius} and poured into plates to obtain approximately \SI{1}{\milli m} thick films. The chitosan layer was prepared by dissolving 1 wt \%  of chitosan (low molecular weight, Sigma Aldrich) powder in \SI{40}{\milli l} DI water that contained 2 wt \% acetic acid. This solution was poured into a silicon mold and placed in the oven for 12 hours at \SI{50}{\degreeCelsius} for the evaporation of water and acetic acid. The film was detached from the silicon mold, cut into the required sizes and washed with an ethanol solution. The final thickness of the chitosan film was in the range of \SIrange{0.2}{0.5}{\milli m}.

For the sensors for the detection of microorganism growth, \SI{12}{\percent} gelatine (Sigma Aldrich) with \SI{5}{\g/L} casein hydrolysate (Sigma Aldrich) and \SI{3}{\g/L} yeast extract (Biolife) was prepared. The mixture was sterilized by autoclaving at \SI{121}{\degreeCelsius} for \SI{20}{\min}, cooled down to \SI{65}{\degreeCelsius} and poured into plates to obtain approximately \SI{1}{\milli m} thick films.

The edible sealant was prepared by mixing 2.5\% \text{w/w} of agar and 3.75\% \text{w/w} of glycerol, and 93.75\% \text{w/w} of water. The solution was placed on a hot plate at \SI{80}{\degreeCelsius} and continuously stirred until a homogenized solution was obtained. The solution was applied to the FP cavity as a sealant.

The hydrophilized droplet generator (Fluidic 440, Chipshop) was used to produce highly monodisperse chlorophyll-doped oil droplets. The chip comprised eight different single cross channels (flow focusing geometry) with nozzle sizes \SIrange{50}{80}{\micro m}. Each channel contained two inlets for the oil and water phase and one outlet for the collection of droplets. Elveflow OB1 pressure controller was used to control the pressure at both inlets precisely. Droplets of \SI{95}{\micro m} diameter were produced by using a single cross channel with a nozzle size of \SI{80}{\micro m} by setting the pressure to \SI{60}{\milli bar} at the chlorophyll-doped oil inlet and \SI{25}{\milli bar} at the water-surfactant inlet. By varying the pressure at one or both inlets and using additional channels with different nozzle sizes, we produced 14 different sizes/samples of monodispersed chlorophyll-doped oil droplets.

The samples were observed with an inverted microscope (Nikon Ti2) through a $20\times$/0.45 NA objective. To attain lasing, a tunable nanosecond pulsed laser (Opotek, Opolette 355) at a repetition rate of \SI{20}{\hertz} was used. The pump wavelength was chosen near the maximum absorption of each dye (Table 1). The resulting fluorescent light or lasing signals were captured by a high-resolution spectrometer (Andor Shamrock SR-500i, Newton) and a digital camera (Andor Zyla) was used for imaging.

\section*{Acknowledgments}

This project has received funding from the European Research Council (ERC) under the European Union’s Horizon 2020 research and innovation programme (grant agreement No. 851143), from the European Union's Horizon 2020 research and innovation programme under the Marie Sklodowska-Curie (grant agreement No. 956265), from Human Frontier Science Program (RGY0068/2020) and from the Slovenian Research And Innovation Agency (ARIS) (P1-0099 and N1-0362). The authors thank Ana Krišelj for the help in preparing the samples and Rok Štanc and Uroš Tkalec for the help with the initial experiments with microfluidics.

\section*{Conflict of Interest}
The authors declare no conflict of interest.

\section*{Author Contributions}
A.R.A. performed most experimental work and analysis of the results. M.M. performed some of the initial experiments and analysis. G.M. and D.N.B. provided the chitosan samples. M.H. conceived the original idea and designed and supervised the study. A.R.A., M.M. and M.H. wrote the manuscript. All authors approved the final version of the paper.

\section*{Data Availability Statement}
The data that support the findings of this study are available from the corresponding author upon reasonable request.

\clearpage

\setcounter{figure}{0}
\setcounter{table}{0}

\section*{Supplementary Information}

\begin{table}[h!]
\centering
\renewcommand{\tablename}{Supplementary Table}
\begin{tabular}{|c| c|} 
\hline
 \textbf{Dye} & \textbf{Solvent}\\ 
 \hline
Coumarin 6 & vegetable oil \\ 
 \hline
NADH &  ethanol  \\
 \hline
 Quinine &  water  \\
 \hline
Sinensetin & dichloromethane   \\
 \hline
 Nor bixin (annatto) & water  \\ 
 \hline
 Chlorophyllin & water \\
 \hline
Curcumin & water \\
 \hline
 E132 Blue & water \\
 \hline
 E102 Yellow & water \\
 \hline
 E122 Red  & water \\
 \hline
 Saffron  & water \\
 \hline
\end{tabular}
\caption{List of the edible dyes that we investigated as a gain medium for different laser types, but did not produce any lasing. 
}
\label{table:S1}
\end{table}

\clearpage

\begin{table}[h!]
\centering
\renewcommand{\tablename}{Supplementary Table}
\begin{tabular}{|c| c|} 
 \hline
 \textbf{Dye} & \textbf{Resonator cavity}\\ 
 \hline
5,10,15,20-Tetraphenyl-21H,23H-porphine & vegetable oil  \\ 
 \hline
meso-Tetraphenylporphyrin&  vegetable oil \\
 \hline
\end{tabular}
\caption{List of the edible dyes that showed WGM modes, but no lasing. 
}
\label{tableS2}
\end{table}

\clearpage

\begin{table}[h!]
\centering
\renewcommand{\tablename}{Supplementary Table}
\begin{tabular}{|c| c|c|} 
 \hline
 \textbf{Sample} & \textbf{ Average diameter (\SI{}{\micro m})}& \textbf{Standard deviation (\SI{}{\nano m})}\\ 
 \hline
 
1 & 39.49 & 157 \\ 
 \hline
2 &  42.88 & 108  \\
 \hline
3 & 45.69 & 165 \\
 \hline
4 & 47.24& 173 \\
 \hline
5 & 52.32 & 136 \\
 \hline
6 & 56.39 & 223 \\
 \hline
7 & 62.30 & 140\\
 \hline
8 & 67.76 & 275 \\
 \hline
9 & 70.38 & 264 \\
 \hline
10 & 73.43 & 213 \\
 \hline
11& 81.45 & 201 \\
 \hline
12 & 84.30 & 148 \\
 \hline
13 & 87.49& 168 \\
 \hline
14 & 94.85 & 224 \\
 \hline
\end{tabular}
\caption{Average diameters and diameter standard deviations of the chlorophyll-doped oil droplets produced by the droplet generator chip, which were used for barcoding. 
}
\label{tableS3}
\end{table}

\clearpage

\begin{figure}[htb]
	\centering
        \renewcommand\figurename{Supplementary Figure}
	\includegraphics[width=13cm]{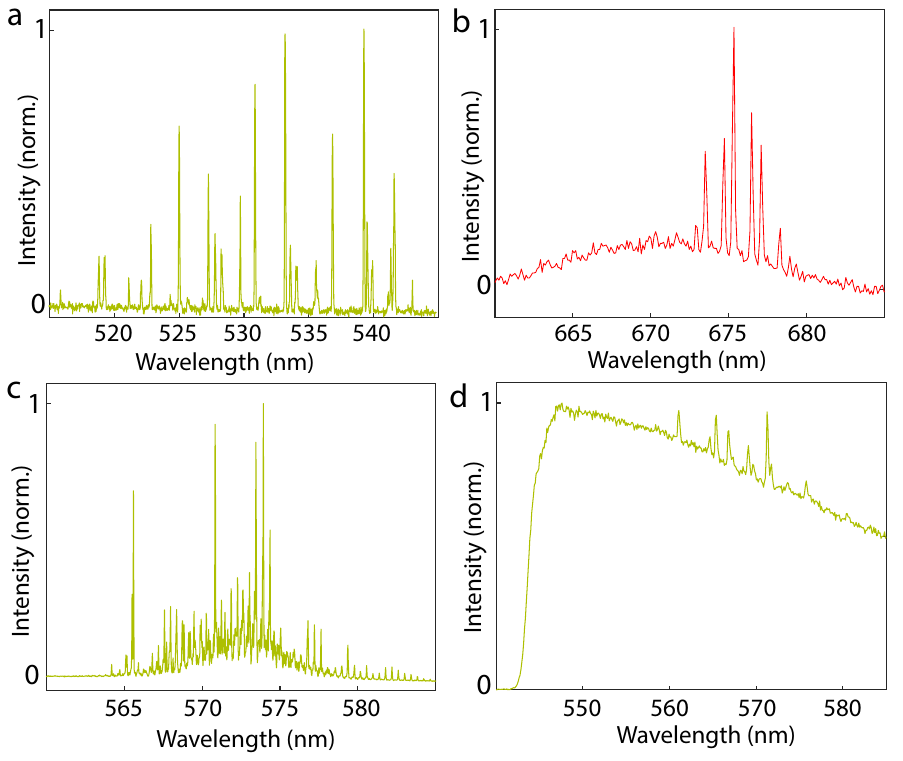}
	\caption{ a) Laser emission spectrum from a bixin-doped oil droplet under the excitation of pulsed laser at \SI{450}{nm}. b) Laser emission spectrum from a non-purified chlorophyll-doped oil droplet under the excitation at \SI{440}{nm}. c) Laser emission spectrum from a riboflavin sodium phosphate-doped water droplet on a spinach leaf under the excitation at \SI{450}{nm}. d) Laser emission spectrum from edible FP cavity filled with riboflavin sodium phosphate solution under the excitation at \SI{450}{nm}.
}
	\label{Supplementary fig:S1}
\end{figure}

\clearpage

\begin{figure}[htb]
	\centering
        \renewcommand\figurename{Supplementary Figure}
	\includegraphics[width=12cm]{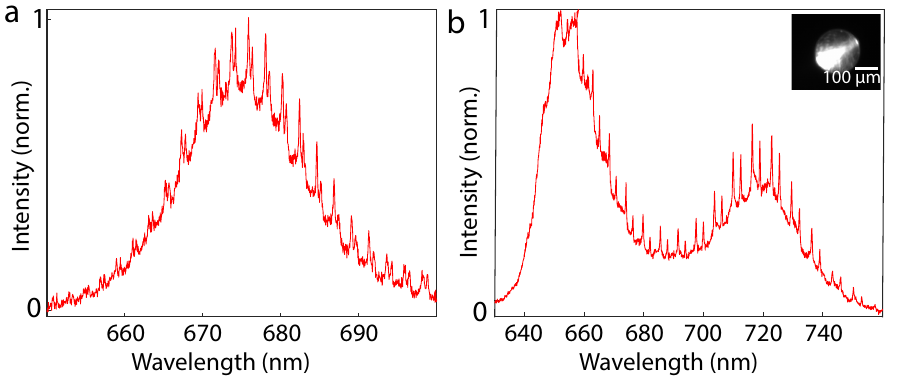}
	\caption{ a) Emission spectrum from a non-purified chlorophyll-doped oil droplet under the excitation of green CW laser (below the lasing threshold). b) Emission spectrum from a porphine-doped oil droplet under the excitation of blue CW laser (below the lasing threshold).
}
	\label{Supplementary fig:S2}
\end{figure}

\clearpage

\begin{figure}[htb]
	\centering
        \renewcommand\figurename{Supplementary Figure}
	\includegraphics[width=12cm]{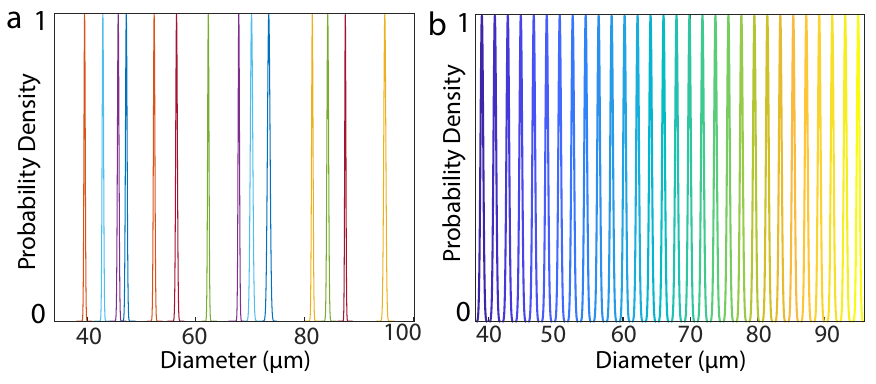}
	\caption{ a) Gaussian curves fitted to the distribution of sizes of the droplets produced by the droplet generator chip. b) In this size range, 30 distinguishable unique sizes that can be fitted without considerable overlap of sizes.
}
	\label{Supplementary fig:S3}
\end{figure}

\clearpage

\begin{figure}[htb]
	\centering
        \renewcommand\figurename{Supplementary Figure}
	\includegraphics[width=15cm]{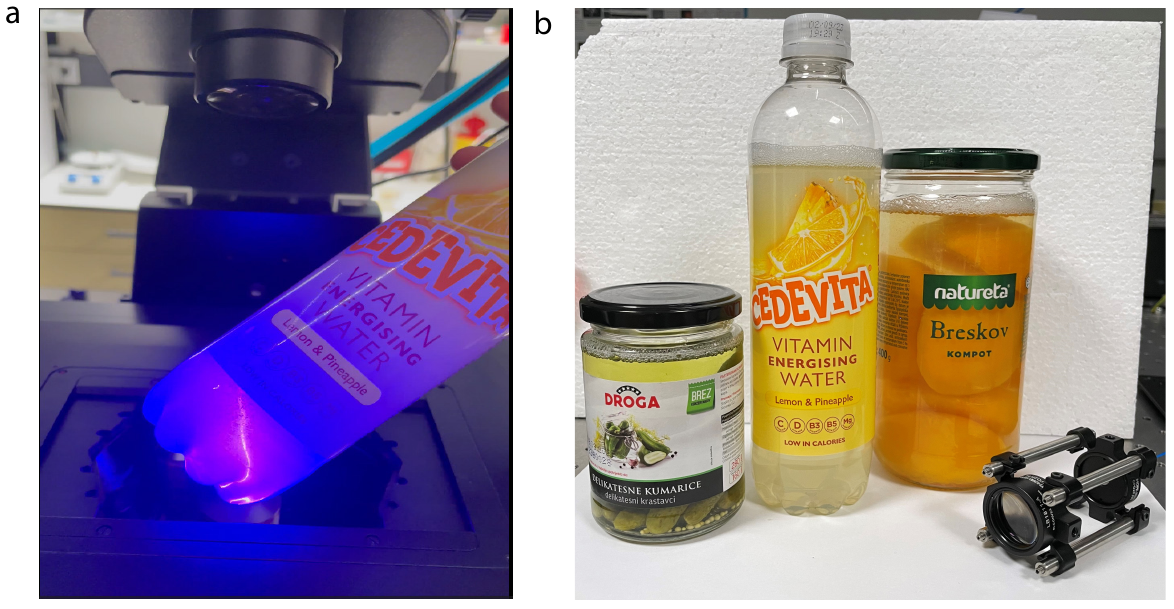}
	\caption{ a) Demonstration of the scanning of the edible barcodes (chlorophyll-doped oil droplets) in juice bottle. b) Barcoding was tested in different products.
}
	\label{Supplementary fig:S4}
\end{figure}

\clearpage

\begin{figure}[htb]
	\centering
        \renewcommand\figurename{Supplementary Figure}
	\includegraphics[width=7cm]{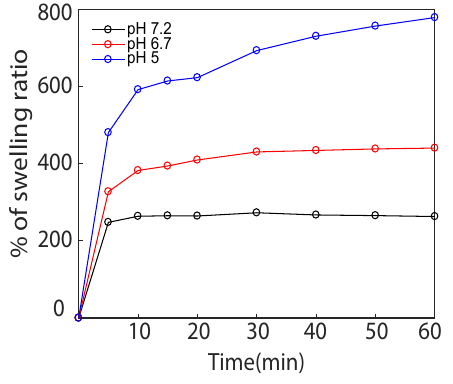}
	\caption{The chitosan film was immersed in different pH solutions, and its weight was measured at various time intervals. The response time was found to be on the order of 10 minutes and the amount of swelling was notably dependent on pH.
}
	\label{Supplementary fig:S5}
\end{figure}

\clearpage

 \typeout{} 
 \bibliography{bib}
 \bibliographystyle{unsrt}
\end{document}